\documentclass[11pt]{article}

\usepackage{amsmath}
\usepackage{amssymb}

\usepackage{graphicx}
\usepackage[tight,footnotesize]{subfigure}
\usepackage{cite}
\usepackage{url}

\usepackage{color} 
\usepackage{soul}
\usepackage{setspace}

\usepackage[labelfont=bf,labelsep=period,justification=raggedright]{caption}

\makeatletter
\renewcommand{\@biblabel}[1]{\quad#1.}
\makeatother

\date{}

\begin{document}

\begin{flushleft}
\begin{spacing}{1.4}
{\LARGE \bf How the Scientific Community Reacts to Newly Submitted Preprints: Article Downloads, Twitter Mentions, and Citations}
\end{spacing}
{\large
{Xin Shuai}$^{1}$, 
{Alberto Pepe}$^{2,\ast}$,
{Johan Bollen}$^{1}$\\}

\vspace*{0.5em}
\textbf {1} School of Informatics and Computing, Indiana University, Bloomington, IN
\\
\textbf{2} Center for Astrophysics, Harvard University, Cambridge, MA
\\
\textbf{$\ast$} Corresponding author: apepe@cfa.harvard.edu
\end{flushleft}

\section*{Abstract}

We analyze the online response to the preprint publication of a cohort of 4,606 scientific articles submitted to the preprint database arXiv.org between October 2010 and May 2011. We study three forms of responses to these preprints: downloads on the arXiv.org site, mentions on the social media site Twitter, and early citations in the scholarly record. We perform two analyses. First, we analyze the delay and time span of article downloads and Twitter mentions following submission, to understand the temporal configuration of these reactions and whether one precedes or follows the other. Second, we run regression and correlation tests to investigate the relationship between Twitter mentions, arXiv downloads and article citations. We find that Twitter mentions and arXiv downloads of scholarly articles follow two distinct temporal patterns of activity, with Twitter mentions having shorter delays and narrower time spans than arXiv downloads. We also find that the volume of Twitter mentions is statistically correlated with arXiv downloads and early citations just months after the publication of a preprint, with a possible bias that favors highly mentioned articles.

\section{Introduction}
The view from the ``ivory tower'' is that scholars make rational, expert decisions on what to publish, what to read and what to cite. In fact, the use of citation statistics to assess scholarly impact is to a large degree premised on the very notion that citation data represent an explicit, objective expression of impact by expert authors \cite{rubin}. Yet, scholarship is increasingly becoming an online process, and social media are becoming an increasingly important part of the online scholarly ecology. As a result, the citation behavior of scholars may be affected by their increasing use of social media. Practices and considerations that go beyond traditional notions of scholarly impact may thus influence what scholars cite. 

\medskip Recent efforts have investigated the effect of the use of social media environments on scholarly practice. For example, some research has looked at how scientists use the microblogging platform Twitter during conferences by analyzing tweets containing conference hashtags \cite{julie,weller:2011}. Other research has explored the ways by which scholars use Twitter and related platforms to cite scientific articles \cite{priem,weller}. More recent work has shown that Twitter article mentions predict future citations \cite{Eysenbach_2011}. This article falls within, and extends, these lines of research by examining the temporal relations between quantitative measures of readership, Twitter mentions, and subsequent citations for a cohort of scientific preprints.

\medskip We study how the scientific community and the public at large respond to a cohort of preprints that were submitted to the arXiv database (\url{http://arxiv.org}), a service managed by Cornell University Library, which has become the premier pre-print publishing platform in physics, computer science, astronomy, and related domains. We examine the relations between three types of responses to the submissions of this cohort of pre-prints, namely the number of Twitter posts (tweets) that specifically mention these pre-prints, downloads of these pre-prints from the arXiv.org web site, and the number of early citations that the 70 most Twitter-mentioned preprints in our cohort received after their submission. In each case, we measure total volume of responses, as well as the delay and span of their temporal distribution. We perform a comparative analysis of how these indicators are related to each other, both in magnitude and time.

\medskip Our results indicate that download and social media responses follow distinct temporal patterns. Moreover, we observe a statistically significant correlation between social media mentions and download and citation count. These results are highly relevant to recent investigations of scholarly impact based on social media data \cite{jason,altmetrics} as well as to more traditional efforts to enhance the assessment of scholarly impact from usage data \cite{clickstream,jcdl,brody,usagebibliometrics}.

\section{Data and study overview}

\subsection{Data collection}
\begin{figure}[h!]
	\begin{center}
	\includegraphics[width=15cm]{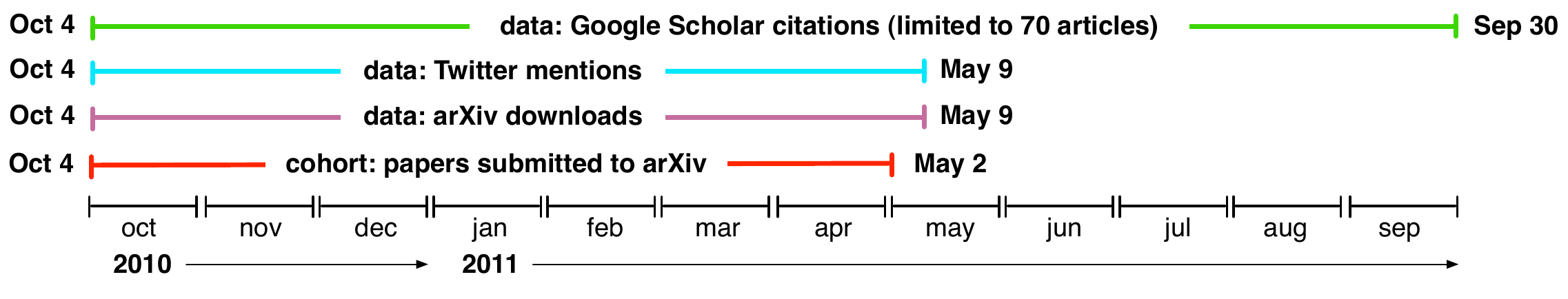}
	\caption{\label{fig:TAC_timeline}  Timeline of data collection. Our cohort consists of all papers submitted to arXiv between October 4, 2010 and May 2, 2011. Weekly article downloads and daily Twitter mentions were recorded after the article's submission date, up to May 9, 2011. Early citation counts for each article were manually recorded from Google Scholar on September 30th, 2011.}
	\end{center}
\end{figure}

Our analysis is based on a corpus of 4,606 scientific articles submitted to the preprint database arXiv between October 4, 2010 and May 2, 2011. For each article in this cohort, we gathered information about their downloads from the arXiv server weekly download logs, their daily number of mentions on Twitter using a large-scale collection of Twitter data collected over that period, and their early citations in the scholarly record from Google Scholar. Table \ref{data_specs} summarizes the discussed data collection and Figure \ref{fig:TAC_timeline} provides an overview of the data collection timelines.

\begin{table}[h!]
\begin{center}
\begin{tabular}{||c|ccc||}
\hline\hline
				&	N				&	articles	&	time period							\\\hline
arXiv downloads	&	2,904,816			&	4,606	&	October 4, 2010 to May 9, 2011			\\
Twitter mentions	&	5,752		&	4,415	&	October 4, 2010 to May 9, 2011			\\
early citations		&	431				&	70		&	October 4, 2010 to September 30, 2011		\\\hline
\end{tabular}
\end{center}
\caption{\label{data_specs} Overview of data collected for a cohort of 4,606 articles submitted to the preprint database arXiv between October 4, 2010 and May 2, 2011.}
\end{table}

\medskip The datasets employed in this study are:
\begin{itemize}
\item \textbf{ArXiv downloads}: For each article in the aforementioned cohort we retrieved their weekly download numbers from the arXiv logs for the period from October 4, 2010 to May 9, 2011. A total of 2,904,816 downloads were recorded for 4,606 articles. 
\item \textbf{Twitter mentions}: Our collection of tweets is based on the Gardenhose, a data feed that returns a randomly sampled 10\% of all daily tweets.  A Twitter mention of  arXiv article was deemed to have occurred when a tweet contained an explicit or shortened link to an arXiv paper (see ``Materials'' appendix for more details). Between October 4, 2010 and May 9, 2011 we scanned 1,959,654,862 tweets in which 4,415 articles out of 4,606 in our cohort were mentioned at least once, i.e. approximately 95\% of the cohort. Such a wide coverage of arXiv articles is mostly due to specialized bot accounts which post arXiv submissions daily. The volume of Twitter mentions of arXiv papers was very small compared to the total volume of tweets in period, with only 5,752 tweets containing mentions of papers in the arXiv corpus. We found that 2,800 out of 5,752 tweets are from non-bot accounts. After filtering out all tweets posted by bot accounts, we retain 1,710 arXiv articles out of 4,415 that are mentioned on Twitter by non-bot accounts. Including or excluding bot mentions, the distribution of number of tweets over all papers was very skewed; most papers were mentioned only once, but one paper in the corpus was mentioned as much as 113 times.
\item \textbf{Early citations}: We manually retrieved citation counts from Google Scholar for the 70 most Twitter-mentioned articles in our cohort. Citation counts were retrieved on September 30, 2011 and date back to the initial submission date in arXiv. All 70 articles combined were cited a total of 431 times at that point. The most cited article in the corpus was cited 62 times whereas most articles received hardly any citations.
\end{itemize}

\medskip By the nature of our research topic, we are particularly focused on \textit{early} responses to preprint submissions, i.e., immediate, swift reactions in the form of downloads, Twitter mentions, and citations. Therefore, we record download statistics and Twitter mention data only one week over the submission period itself (up to May 9, 2011). As for citation data, we are aware that citations take years to accrue.  We do not explore here long-term citation effects, but only the early, immediate response to pre-print submission in the form of citations in the scholarly record. Our citation data pertains to a time period that spans from 5 months to 1 year: it is a fraction of the expected amount of ``maturation time'' for citation analysis. Citation data must therefore be considered to reflect ``early citations'' only, not total potential citations.

\subsection{Definitions: delay and time span.}
Twitter mentions and arXiv downloads may follow particular temporal patterns. For example, for some articles downloads and mentions may take weeks to slowly increase after submission, whereas for other articles downloads may increase very swiftly after submission to wane very shortly thereafter. The total number of downloads and mentions is orthogonal to these temporal effects, and could be different in either case.

\medskip The two parameters that we use to describe the temporal distributions of arXiv downloads and Twitter mentions are \emph{delay} and the time \emph{span}, which we define as follows. Let $t_0 \in \mathbb{N}^+$ be the date of submission for article $a_i$. We represent both arXiv downloads and Twitter mentions for article $a_i$ as the time series $T$, the value of which at time $t$ is given by the function $\mathbb{T}(a_i,t) \in \mathbb{N}^+$. We then define the time of the first, maximum, and last arXiv download of article $a_i$ as $\mathbb{T}_{\text{first}}(a_i)$,  $\mathbb{T}_{\text{max}}(a_i)$, and $\mathbb{T}_{\text{last}}(a_i)$ respectively:

\[
	\begin{array}{l}
	\mathbb{T}_{\text{first}}(a_i) = \min\{t:\mathbb{T}(a_i,t) > 0\}\\
	\mathbb{T}_{\text{max}}(a_i) = t: \max(\mathbb{T}(a_i,t) \\
	\mathbb{T}_{\text{last}}(a_i) = \max\{t:\mathbb{T}(a_i,t) > 0\}\\
	\end{array}
\]

The delay, $\Theta(a_i)$, and span, $\Delta(a_i)$, of the temporal distribution of arXiv downloads for article $a_i$ will then be defined as:

\[
	\begin{array}{l}
		\Theta(a_i) = \mathbb{T}_{\text{last}}(a_i) - t_0 \\
		\Delta(a_i)=\mathbb{T}_{\text{last}}(a_i) - \mathbb{T}_{\text{first}}(a_i)
	\end{array}
\]

To distinguish between the delay and span of arXiv downloads and twitter mentions, we simply denote $\Theta_{\text{ax}}(a_i)$, $\Delta_{\text{ax}}(a_i)$, $\Theta_{\text{tw}}(a_i)$, $\Delta_{\text{tw}}(a_i)$ respectively which are defined according to the above provided definitions.

\begin{figure}[h!]
	\begin{center}
	\includegraphics[width=0.6\textwidth]{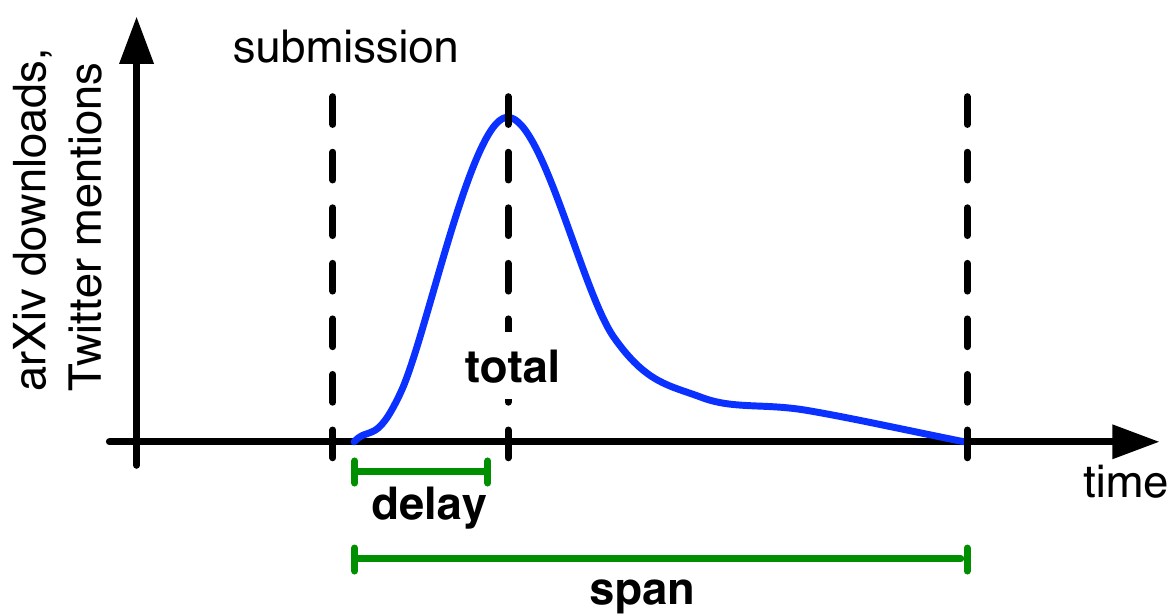}
	\caption{\label{fig:span_delay}  Span and delay of temporal distribution of arXiv downloads or Twitter mentions over time expressed in terms of time passed between submission of article and peak and time passed between first and last event, respectively.}
	\end{center}
\end{figure}

\medskip As shown in Figure \ref{fig:span_delay}, the delay is thus measured as the time difference between the date of a preprint submission and a subsequent spike in Twitter mentions (the day in which an article receives the highest volume of related tweets) or arXiv downloads (the day in which it receives the highest volume of downloads). The time span is the temporal ``duration'' of the response, measured as the time lag between the first and the last Twitter mention or download of the article in question.

\begin{figure}[h!]
	\centering
	\includegraphics[width=1\textwidth]{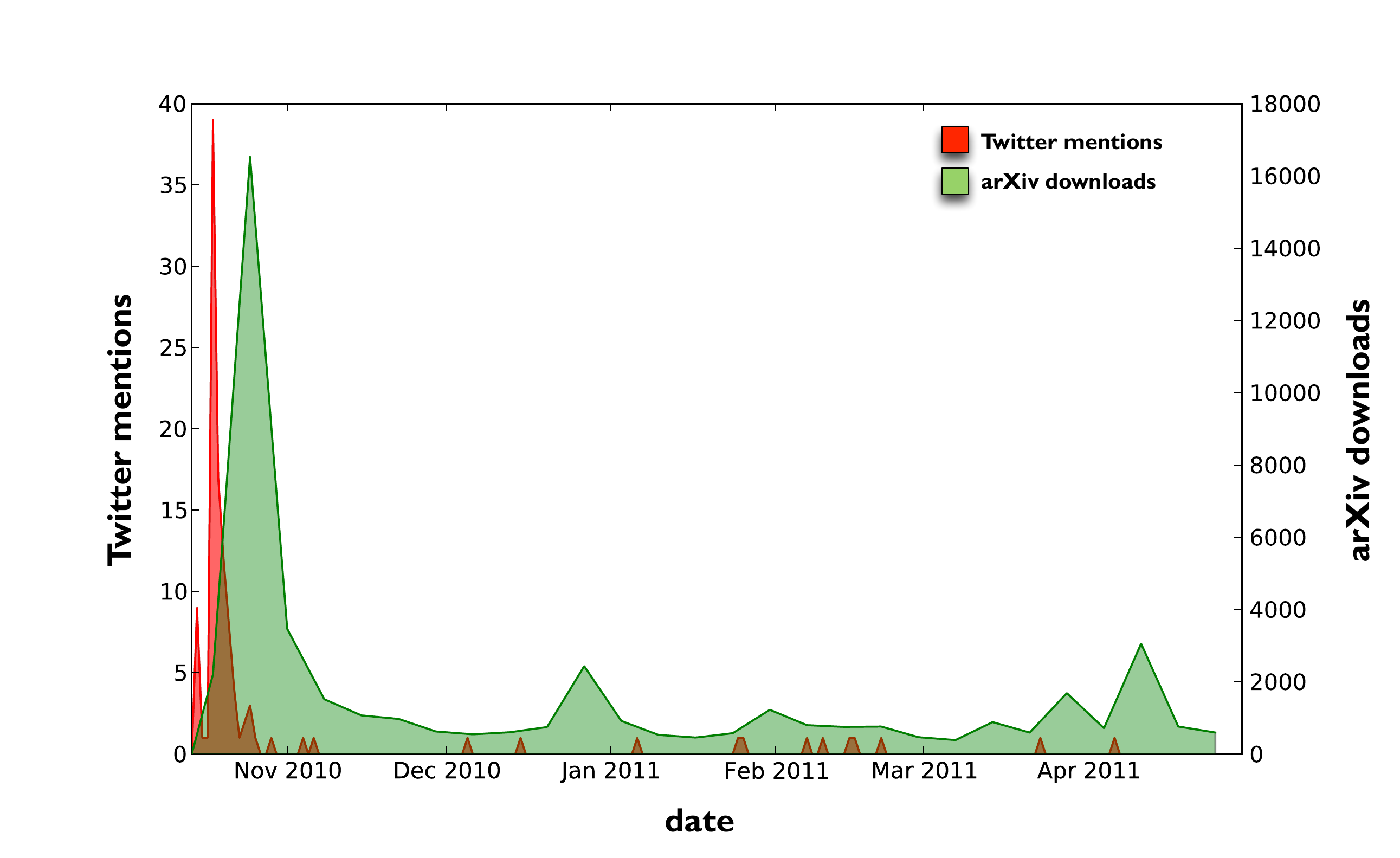}
	\caption{\label{fig:spikes} Response dynamics (Twitter mentions and arXiv downloads) for a selected arXiv preprint. As shown, for this particular example, Twitter mentions spikes shortly after submission and publication, and wane quickly with very mentions after the initial burst. ArXiv downloads peak shortly afterwards but continue to exhibit significant activity many weeks later.}
\end{figure}

\medskip To illustrate delay and span, we examine in detail the response dynamics for an article in the corpus, in Figure \ref{fig:spikes}. The article in question was submitted to arXiv on October 14, 2010. Time runs horizontally from left to right. Downloads and Twitter mentions are charted over time (weekly for downloads, daily for mentions). As Figure \ref{fig:spikes} shows, the Twitter response to submission occurs within a day, reaching a peak of nearly 40 daily mentions within several days, and then slowly dies out over the course of the following week. The peak of arXiv downloads, with over 16,000 weekly downloads, occurs a couple of weeks after submission, and continues to be marked by downloads for months. From a \textit{post hoc, ergo propter hoc} point of view, in this case the Twitter response occurs immediately and nearly exactly before the peak in arXiv reads, suggesting that social media attention may have led to subsequently higher levels of arXiv downloads.

\section{Results and discussion}
In this section, we present three results: descriptive statistics of arXiv downloads and Twitter mentions, a temporal analysis of time span and delay in arXiv downloads and Twitter mention, and a regression analysis between arXiv downloads, Twitter mentions, and early citations. For the descriptive statistics, we keep all 5,752 tweets and 4,415 articles mentioned on Twitter, since we want to show a full picture of our data. For the subsequent temporal and regression analysis we only focus on the 2,800 tweets and 1,710 arXiv articles mentioned by non-bot accounts to avoid spurious effects introduced by automated bot accounts.

\subsection{Domain-level descriptive statistics}
\begin{figure}[h!]
	\centering
	\subfigure[]{
	    \label{fig:rank_subs1}
		\includegraphics[scale=0.35]{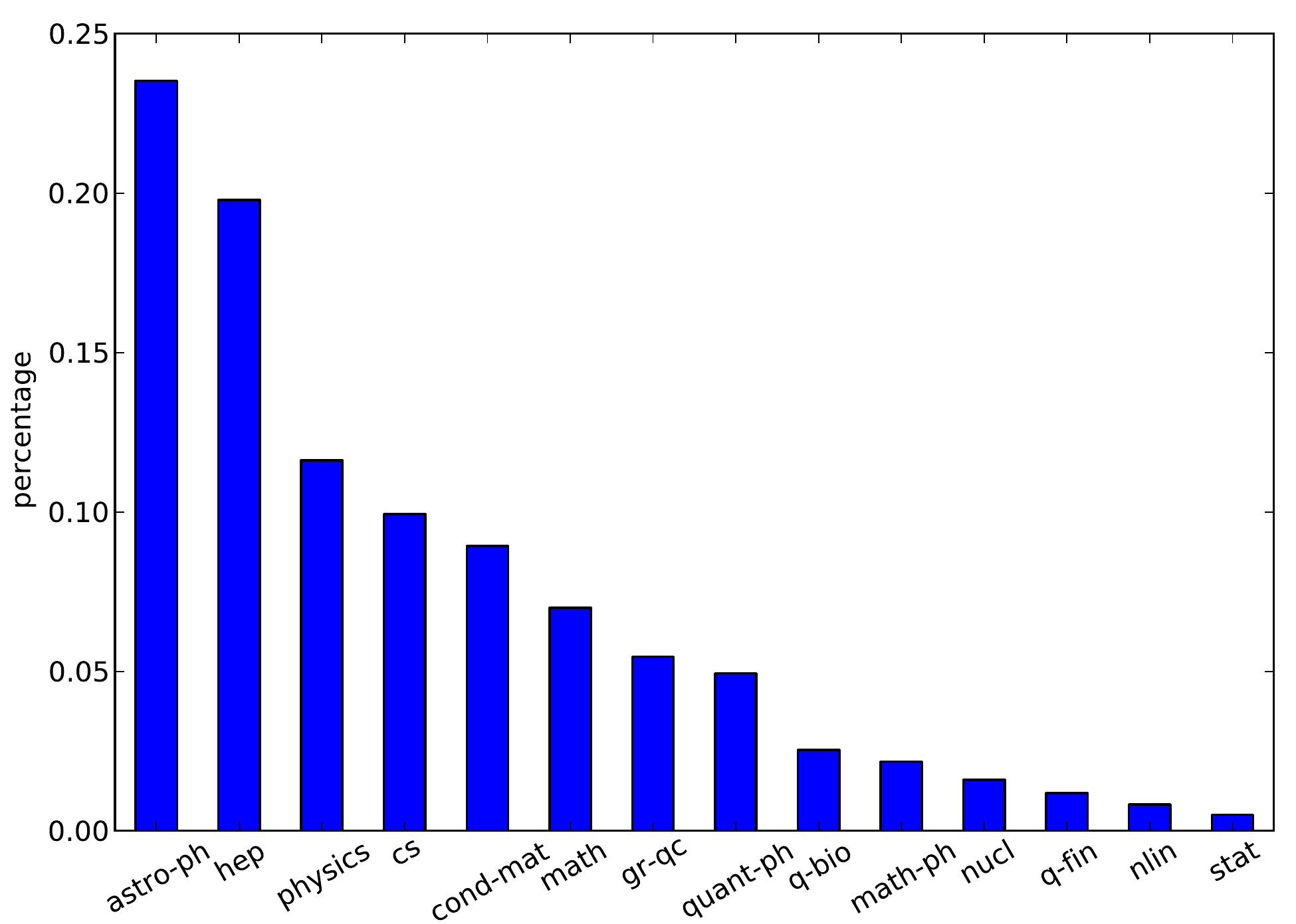}
	}
	\subfigure[]{
	    \label{fig:rank_subs2}
		\includegraphics[scale=0.35]{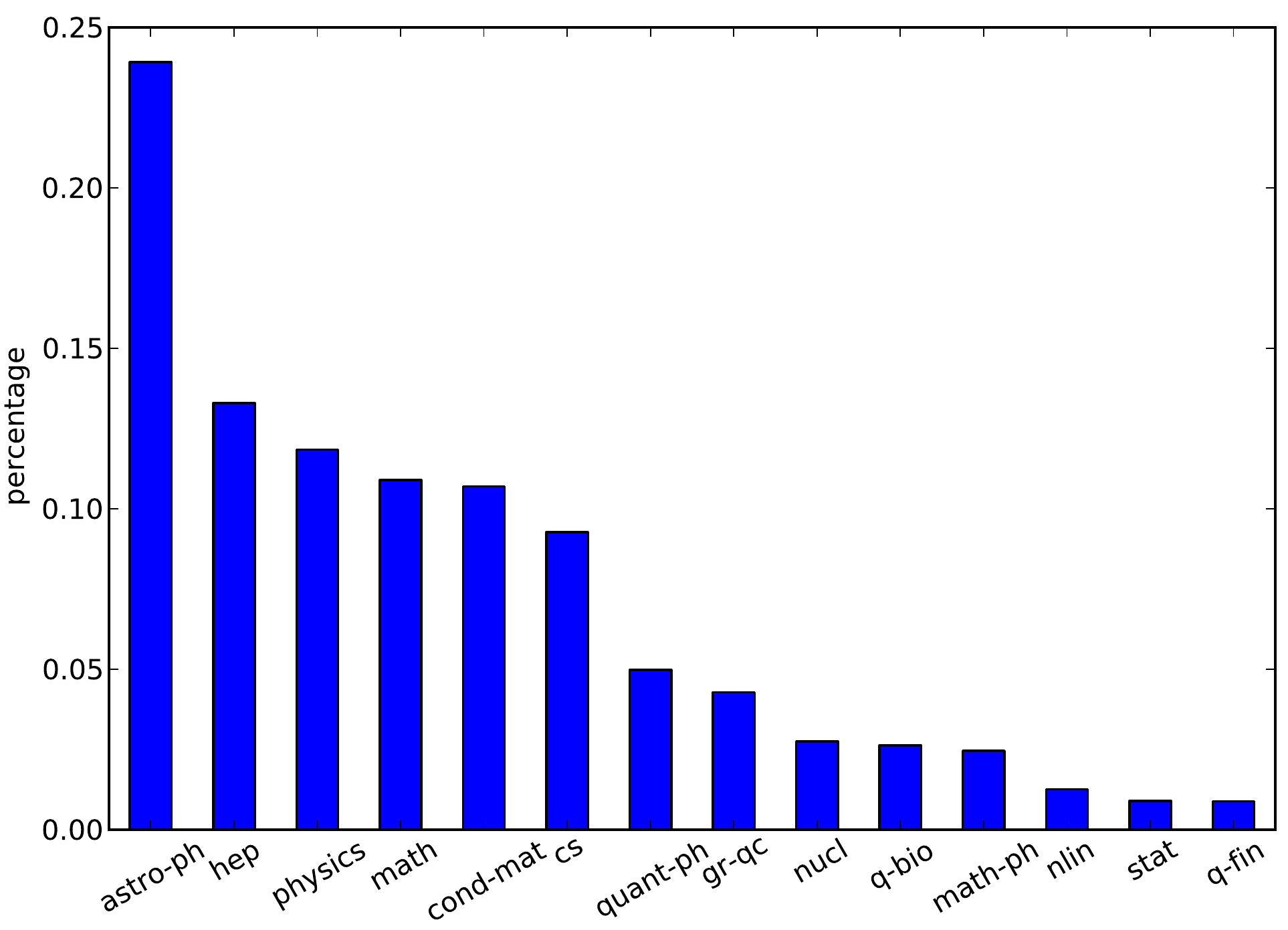}
	}
	\subfigure[]{
	   \label{fig:arxiv_temporal}
	   	\includegraphics[scale=0.35]{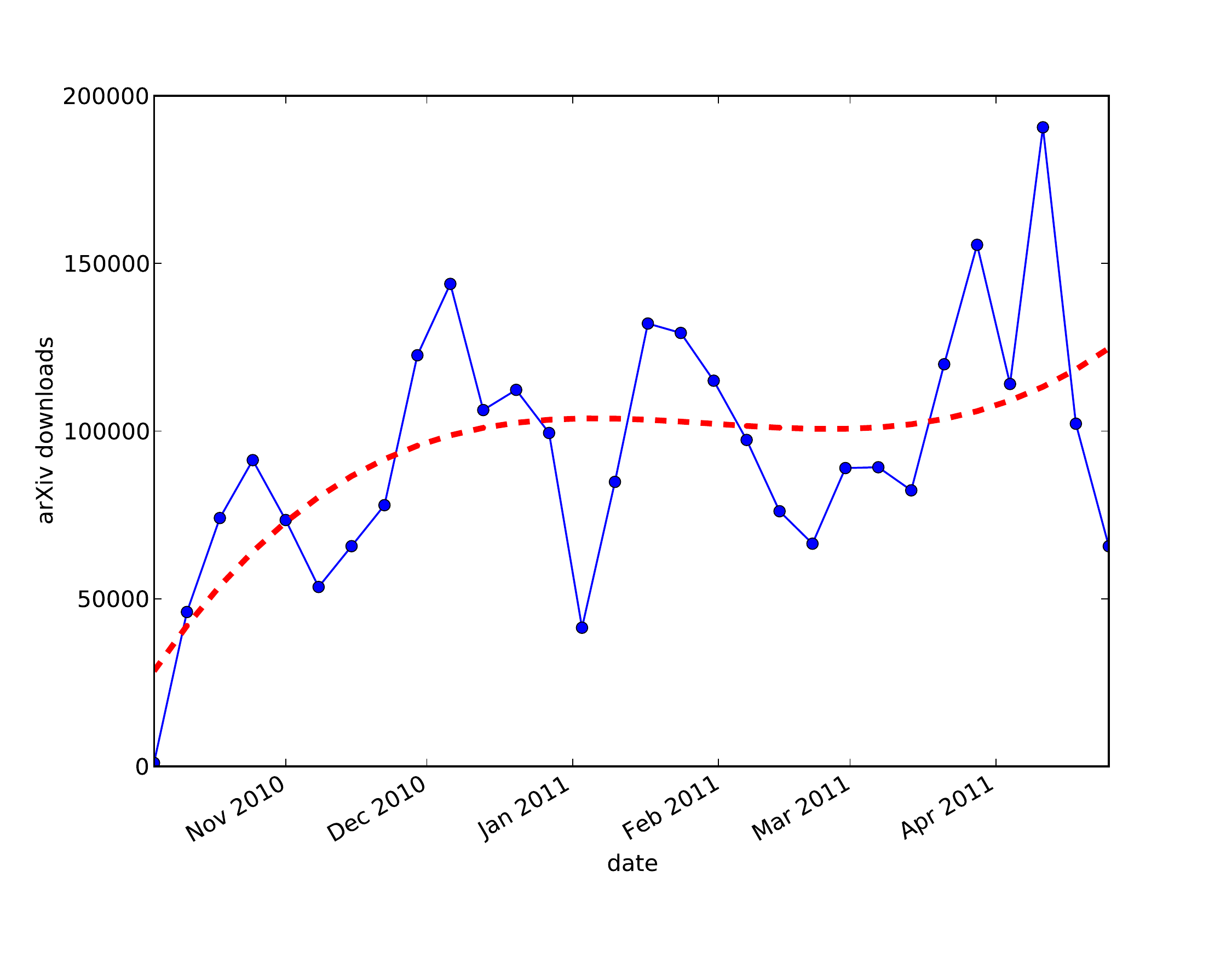}
	}
	\subfigure[]{
	   \label{fig:total_tweets_temporal}
	   	\includegraphics[scale=0.35]{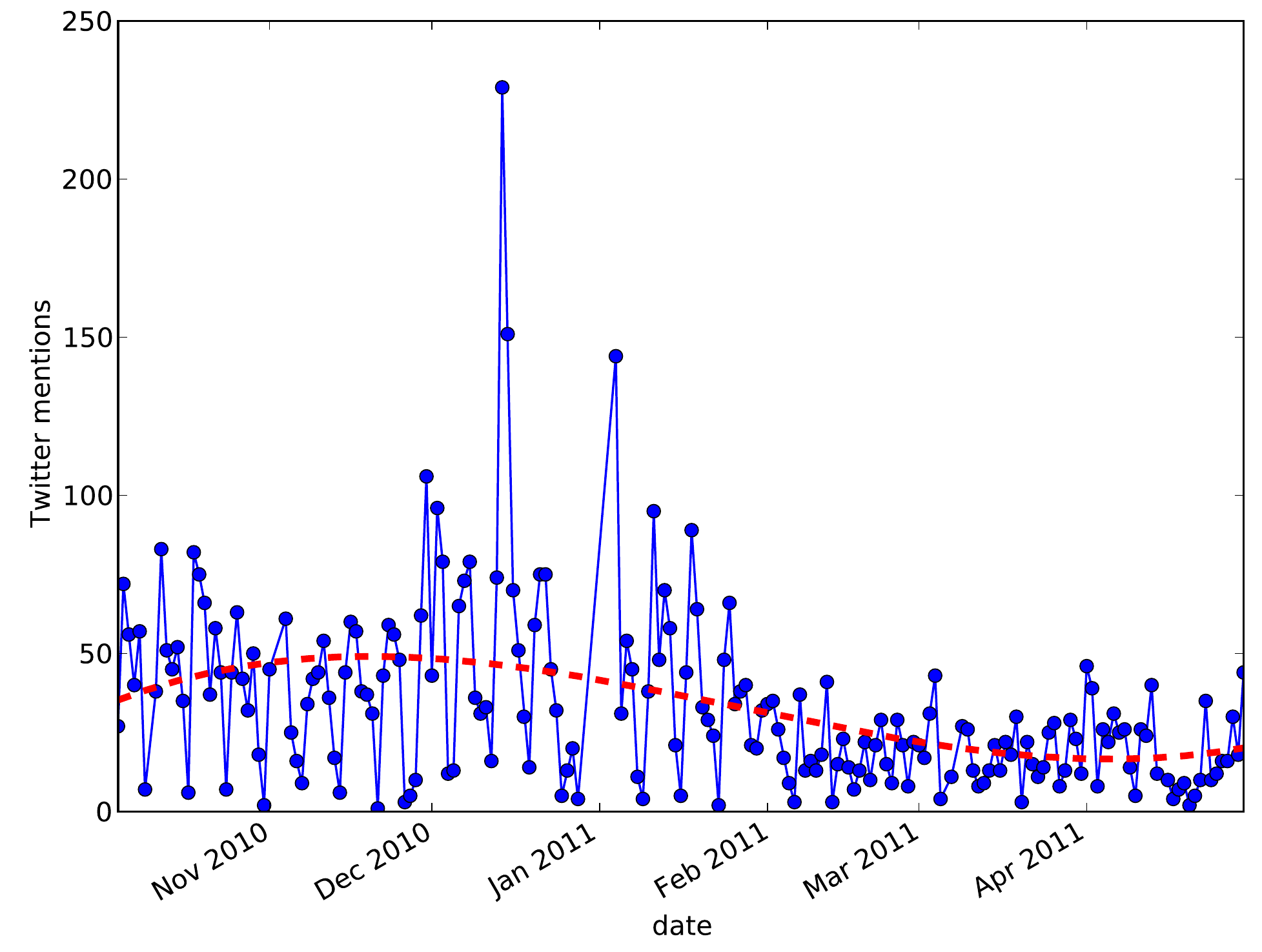}
	}
	\caption{\label{fig:descriptive} (a) Barplot of frequency of subject domains for downloaded paper (rank-ordered), (b) 
	  Barplot of frequency of subject domains for Twitter-mentioned papers (rank-ordered), (c) temporal distribution of total arXiv
	  downloads (weekly), and (d) temporal distribution of total Twitter mentions of arXiv papers in our cohort.}
\end{figure}

Some descriptive statistics about the datasets analyzed in this article are presented in Figure  \ref{fig:descriptive}. The first row of plots in Figure \ref{fig:descriptive} displays the arXiv subject domains of (a) downloaded, and (b) Twitter mentioned papers (by percentage). A full list of the subject domain abbreviations used in these plots is available in the Materials section, Table \ref{arxivabbrv}. We observe a broad and evenly spread distribution of subject domains for downloads and mentions: most papers downloaded and mentioned on Twitter relate to Physics, in particular Astrophysics, High Energy Physics, and Mathematics. The second row of plots in Figure \ref{fig:descriptive} displays the temporal distributions of (c) downloads, and (d) Twitter mentions (the dotted line in both figures is obtained by fitting a 3rd order polynomial function for smoothing). As shown in Figure \ref{fig:arxiv_temporal}, download counts of articles increase over time. This may be partly caused by a cumulative effect: papers that were published earlier have had more time to accumulate reads than papers that were published later. Figure \ref{fig:total_tweets_temporal}, however, shows that the total number of tweets that mention arXiv papers decreases over time.
 
\begin{figure}[h!]
         \centering
	\includegraphics[width=14cm]{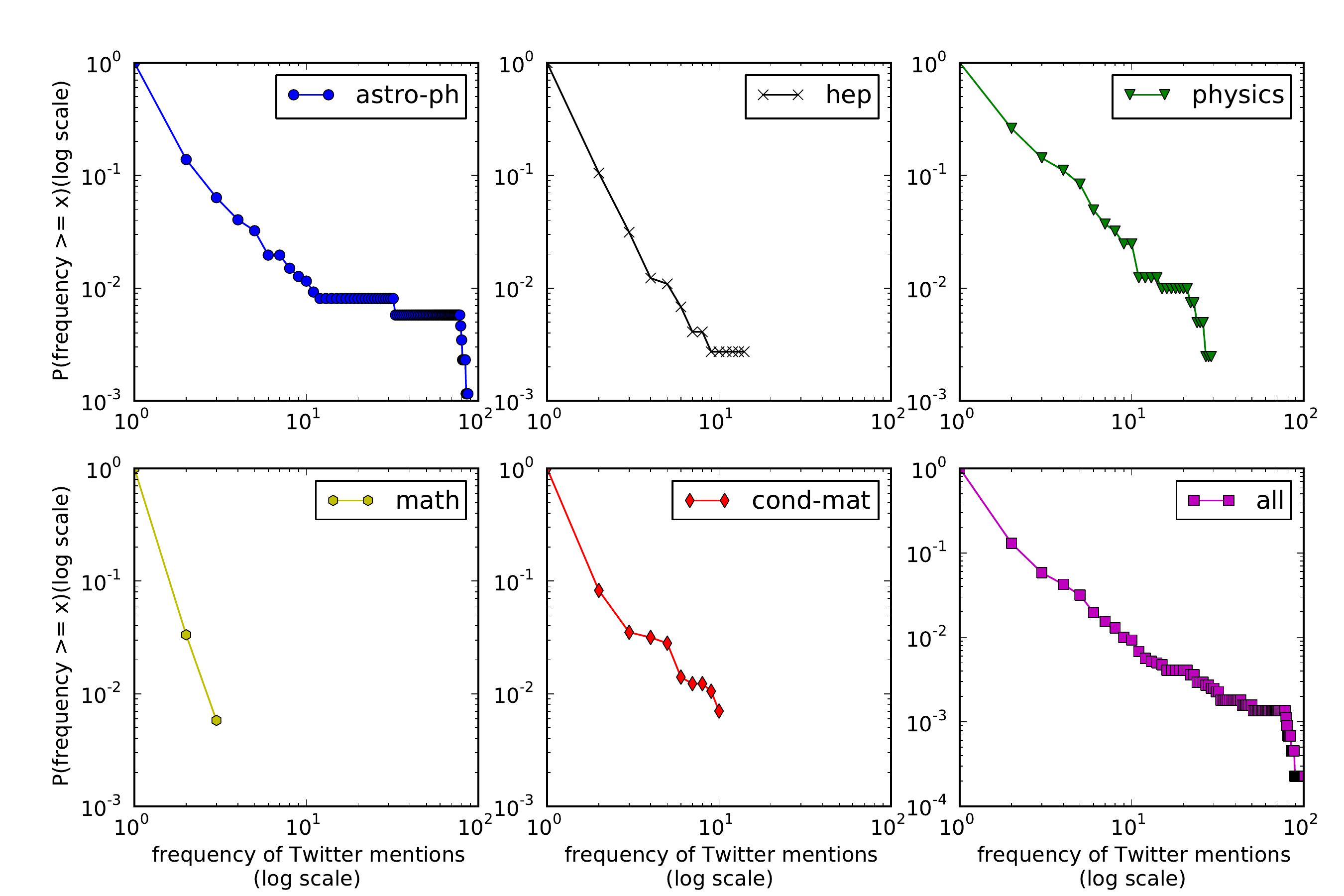}
	\caption{\label{fig:sub_dist}  Complementary Cumulative Distribution Functions (CCDF) of Twitter mentions for all articles in the 5 most frequently observed subjects domains.}
\end{figure}

\medskip In order to better understand how Twitter mentions vary across domain, we show the Complementary Cumulative Distribution Functions (CCDF) of Twitter mentions for all articles in the five most frequently observed subjects domains of Figure \ref{fig:sub_dist}. We find that within each domain few papers receive relatively many mentions whereas the majority receive very few. The frequency-rank distribution is thus strongly skewed towards low values indicating that most articles  receive very few Twitter mentions. Note that we rely on the so-called Twitter Gardenhose, a random sample of about 10\% of all daily tweets, and may thus underestimate the absolute number of Twitter mentions by a factor of 10. (Refer to Materials section for more details).

\subsection{Temporal analysis of delay and span}

\begin{figure}[h!]
  \centering
  \subfigure[]{
	\includegraphics[width=0.7\textwidth]{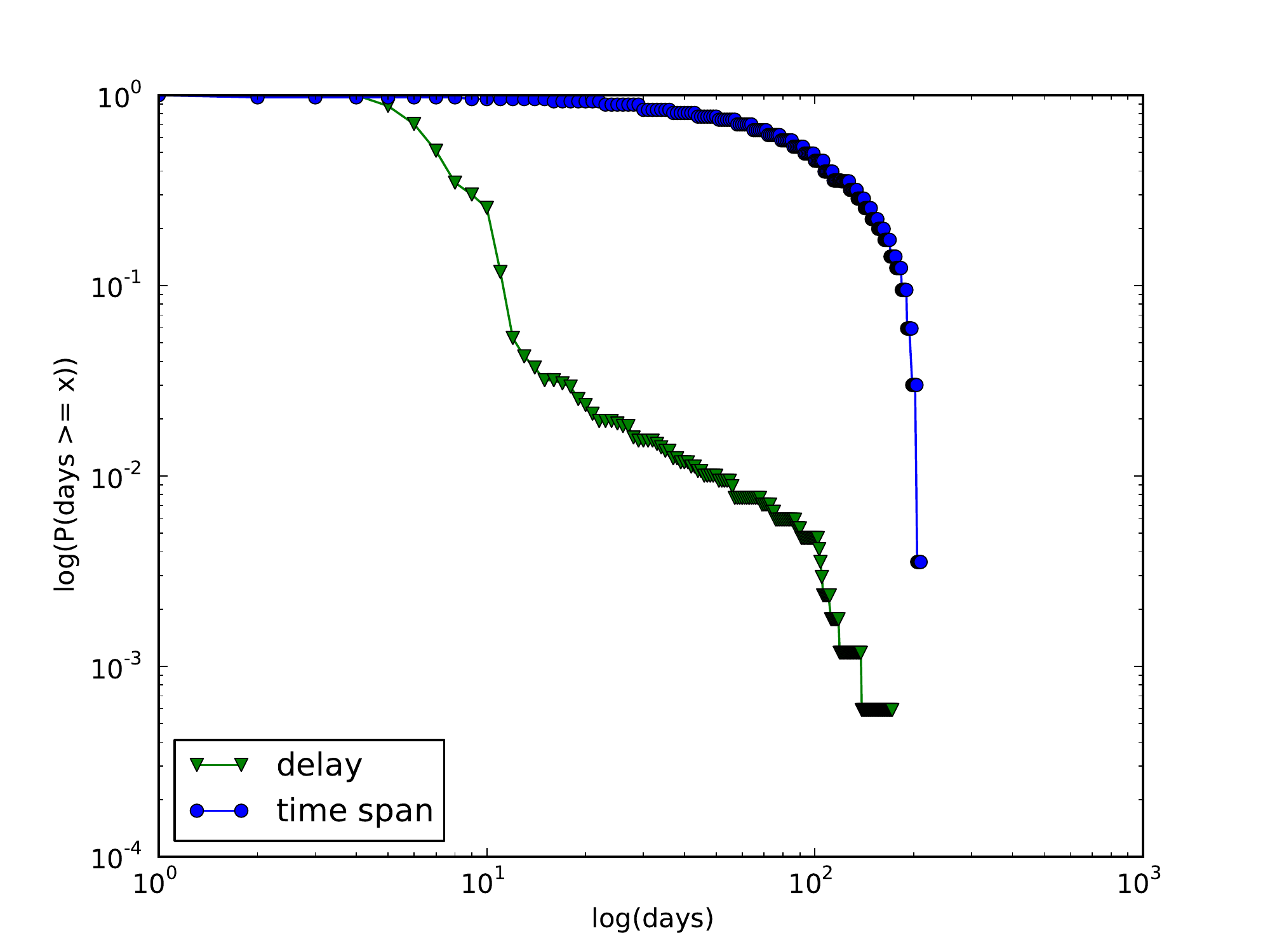}
	\label{fig:delay_span_arxiv}}
 \subfigure[]{
         \includegraphics[width=0.7\textwidth]{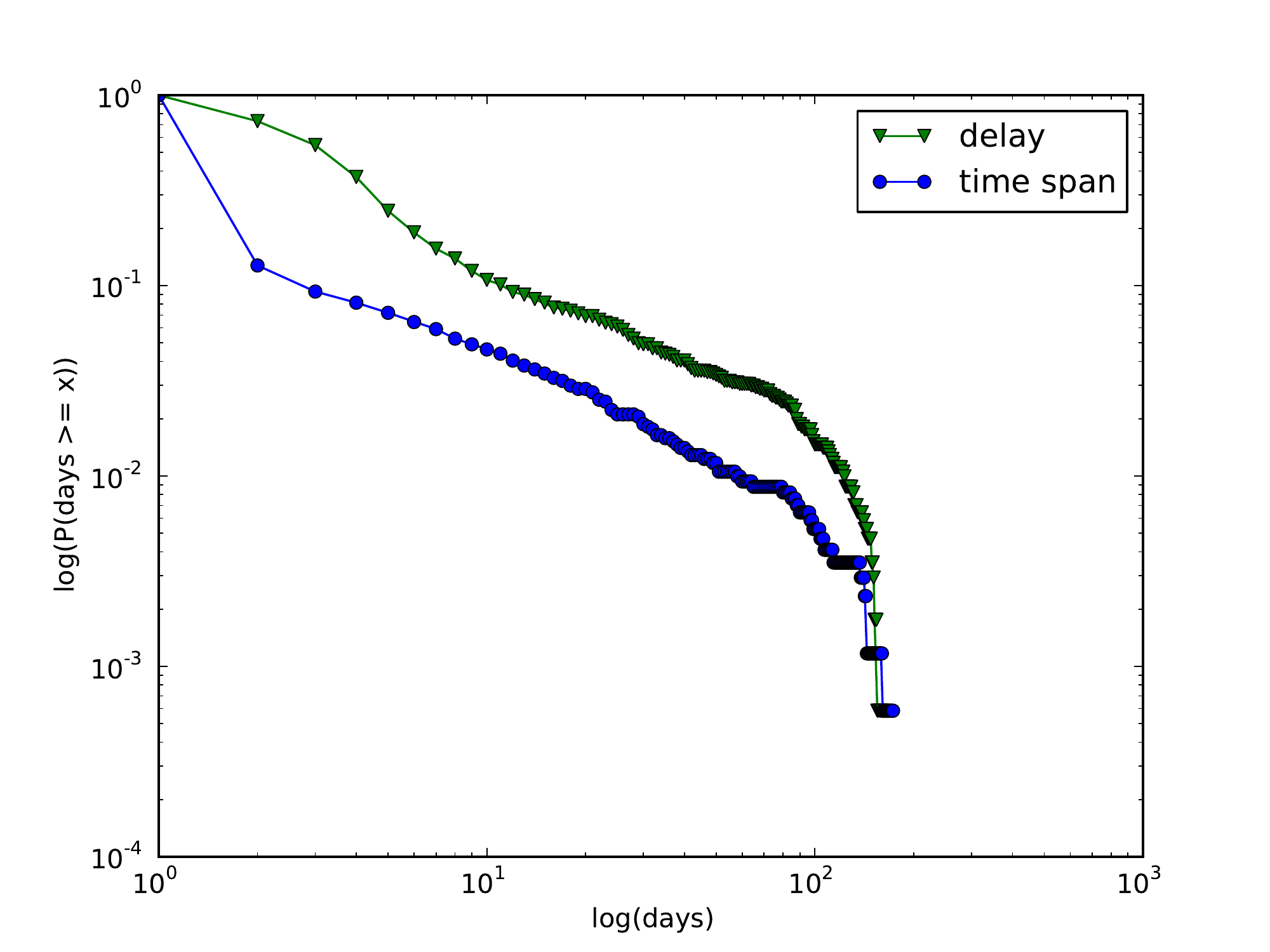}
         \label{fig:delay_span_twitter}}
\caption{\label{fig:delay_span} Distributions of $\log(P(days \ge x))$ for delay and span values in $\log(days)$ for (a) arXiv downloads and (b) Twitter mentions, recorded for all arXiv submissions in our cohort.}
\end{figure}

\medskip In Figure \ref{fig:delay_span}, we plot the distributions of $\Theta_{\text{ax}}(a_i)$, $\Delta_{\text{ax}}(a_i)$, $\Theta_{\text{tw}}(a_i)$, and $\Delta_{\text{tw}}(a_i)$ following article submission. We can see that the distributions of $\Theta_{\text{ax}}(a_i)$, $\Delta_{\text{ax}}(a_i)$, $\Theta_{\text{tw}}(a_i)$, and $\Delta_{\text{tw}}(a_i)$ are highly skewed towards very low values, with very few cases characterized by extensive delays or time spans. In Figure~\ref{fig:delay_span_arxiv}, the distribution of  $\Theta_{\text{ax}}(a_i)$ curve shows that nearly all articles take at least 5 days to reach the peak of arXiv downloads ($x<=4: y=1$), i.e., all articles take more than 4 days to reach peak downloads. In addition, the distribution of $\Delta_{\text{ax}}(a_i)$ curve shows that most of the articles are downloaded persistently for over 100 days ($x<=100: y>0.6$). 

\medskip From Figure \ref{fig:delay_span_twitter}, it emerges that nearly 80\% of the articles in the corpus reach the peak of Twitter mention just one day after they are submitted, as is shown on the distribution of $\Theta_{\text{tw}}(a_i)$ curve ($x=2:y \simeq0.8$). Over 70\% articles reach the peak of Twitter mention within 5 days of submission ($x=5: y < 0.3$). However, the distribution of  $\Delta_{\text{tw}}(a_i)$ curve shows that over 80\% of arXiv.org articles are mentioned one and one day only ($x=2: y<0.2$), i.e., one or multiple tweets about an article are posted within the time range of 24 hours and then are never mentioned again. Overall, compared with arXiv downloads, the Twitter response to scientific articles is typically swift, yet highly ephemeral, a pattern indicative of a process in which the news of a publication is quickly passed around and very little in-depth discussion taking place afterward.

\subsection{Regression between article downloads, Twitter mentions, and citations}
We investigate the degree by which article citations, denoted $C$, can be explained in terms of article-based Twitter mentions, denoted $T$, and arXiv downloads, denoted $A$, by means of a multi-variate linear regression analysis. This analysis is limited to a cohort of the 70 most mentioned articles on Twitter that were submitted to arXiv.org from October 4, 2010 to March 1, 2011 (5 months). This limitation is due to the extent of work involved in manually collecting early citation data as well as to the fact that a cohort of articles submitted earlier in the timeline can provide a fuller coverage of Twitter mentions and arXiv downloads.  For each article, we retrieve the total number of Twitter mentions and arXiv downloads 60 days after submission, and their total number of early citation counts on September 30, 2011 (7 months later after submission of the latest paper).

\medskip Given that each article could have been submitted at any time in a 5 month period, i.e. October 4, 2010 to March 1, 2011, on September 30, 2011 some articles could have had 5 more months than others to accumulate early citations. Therefore the citation counts observed on September 30, 2011 may be biased by the submission date of the article in question. We must therefore include the amount of time that an article has had to accumulate citations since their submission date as an independent variable in our regression models.

\medskip Let $P$ represent the number of days between the submission time of the article and September 30, 2011. We thus define the following multivariate linear regression models:

\begin{equation}
C=\beta_{1}T+\beta_{1}P+\varepsilon
\end{equation} 
\begin{equation}
C=\beta_{1}A+\beta_{2}P+\varepsilon
\end{equation}
\begin{equation}
C=\beta_{1}T+\beta_{2}A+\beta_{3}P+\varepsilon
\end{equation}
where $\beta_{i}$ denotes the corresponding regression coefficient.

\medskip From Table~\ref{tbl:multireg}, we observe that publication period $P$ is certainly a non-neglectable factor to predict the citation counts $C$ but also that Twitter mentions $T$ shows equally significant correlations. Moreover, Twitter mentions seem to be the most significant predictor of citations, compared to arXiv downloads and time since publication. This is not the case for arXiv downloads which, when accounting for Twitter mentions and arXiv downloads, do not exhibit a statistically significant relationship to early citations.

\begin{table}
	\caption{Multi variant linear regression analysis of article citations $C$ vs.~twitter mentions $T$, article arXiv downloads $A$, and time in days elapsed between beginning of our test period and submission of article, $P$. }
	\label{tbl:multireg}
	\centering
	\begin{tabular}{||l||l|l|l||}
	\hline\hline
	model										&	$\beta_1$ (st. error)			&		$\beta_2$ (st. error)			&	$\beta_3$ (st. error)		\\\hline
	$C=\beta_{1}T+\beta_{2}P+\varepsilon_1$			&	$0.150^{***}$ (0.035)		&		$0.044^{**}$(0.019)			&	-					\\
	$C=\beta_{1}A+\beta_{2}P+\varepsilon_2$			&	2e-04$^{***}$ (7e-05)		&		$0.038^{*}$(0.020)			&	-					\\
	$C=\beta_{1}T+\beta_{2}A+\beta_{3}P+\varepsilon_3$ 	&	$0.120^{***}$ (0.040)		&		1e-04(8e-05)				&	$0.041^{**}$(0.019)			\\\hline\hline
	\multicolumn{4} {||l||} {*: p$<$0.1,**: p$<$0.05,***: p$<$0.01,****: p$<$0.001} \\\hline\hline
	\end{tabular}
\end{table}

\begin{figure}[h!]
	\centering
	\subfigure[]{
	   \label{fig:twitter_arxiv_70}
	   	\includegraphics[width=0.31\textwidth]{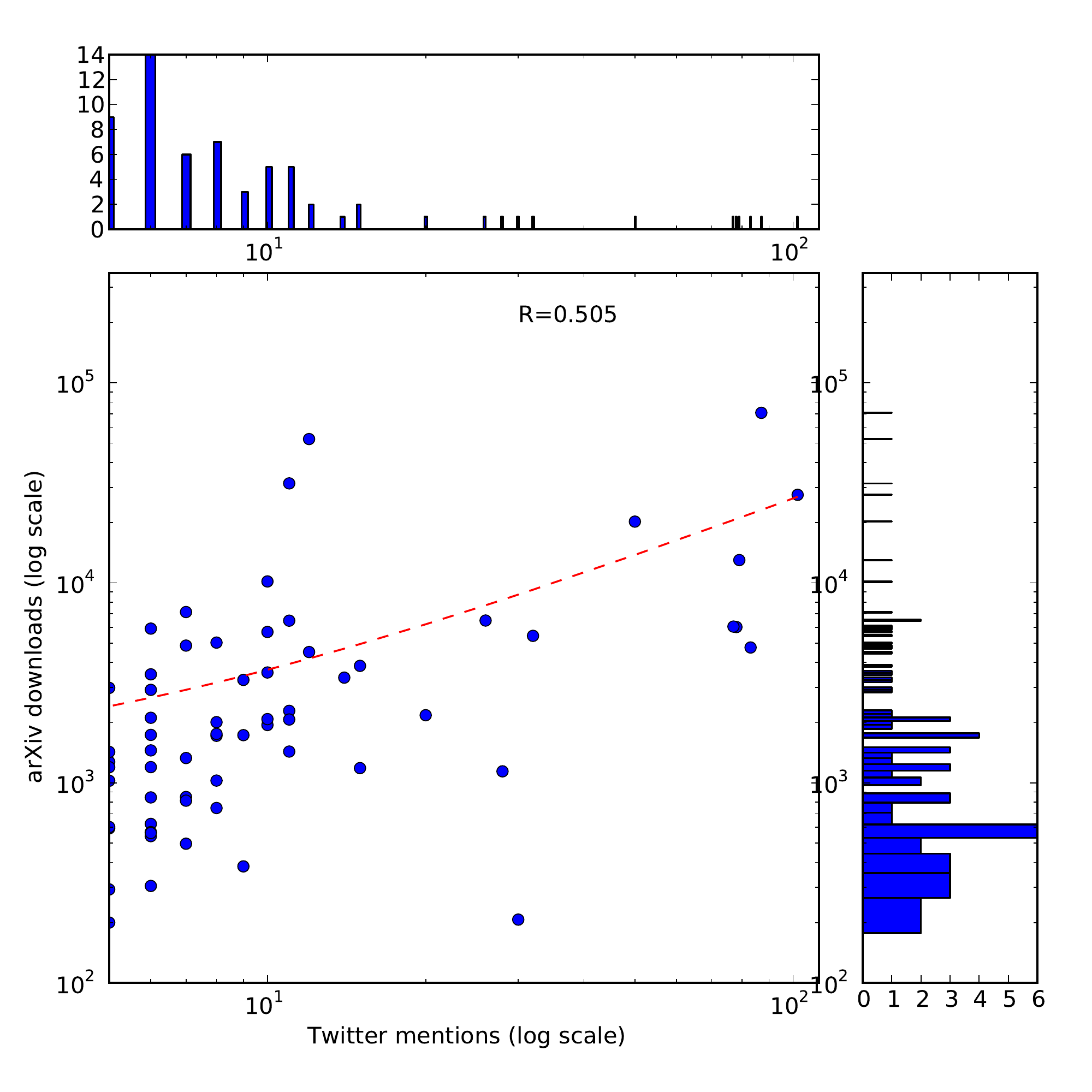}
		\label{fig:tvsa}
	}
	\subfigure[]{
	   \label{fig:twitter_citation_70}
	   	\includegraphics[width=0.31\textwidth]{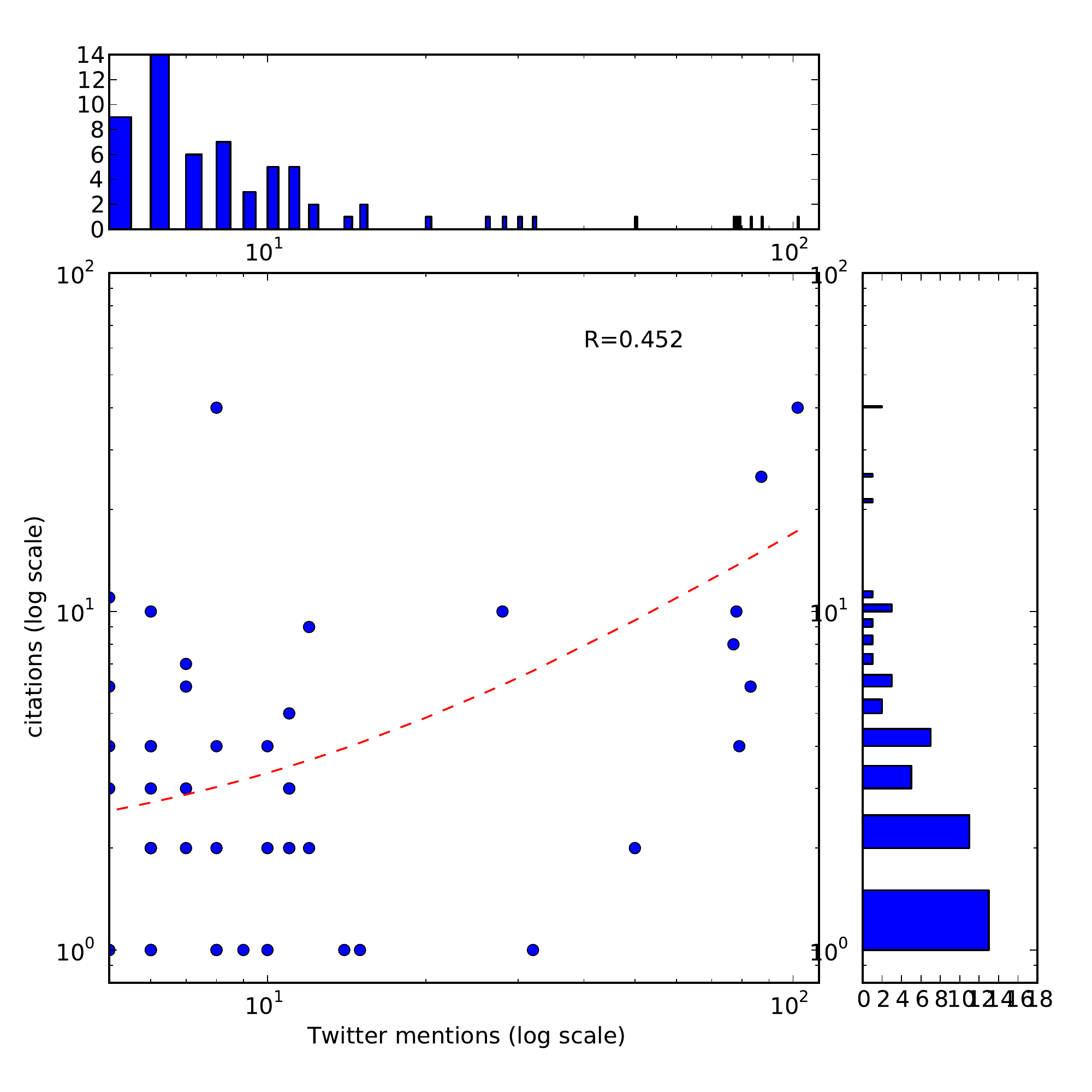}
		\label{fig:tvsc}
	}
	\subfigure[]{
	   \label{fig:arxiv_citation_70}
	   	\includegraphics[width=0.31\textwidth]{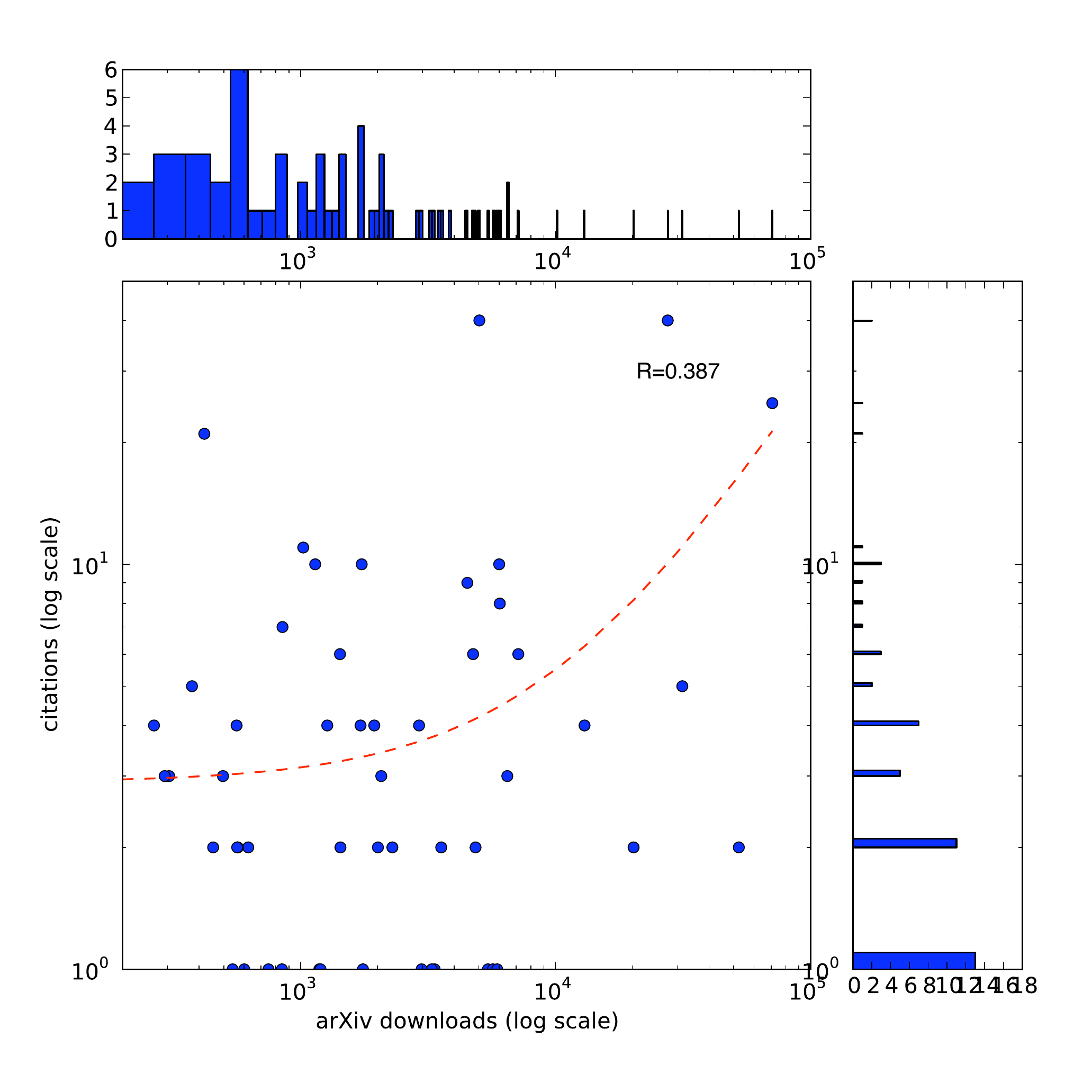}
		\label{fig:avsc}
	}
	  \caption{\label{fig:scatterplots}  Log-log scatter plots of (a) Twitter mentions vs. arXiv downloads, (b) Twitter mentions vs. citations and (c) arXiv downloads vs. citations for 70 most mentioned articles on Twitter indicate statistically significant correlations. Marginal densities of distributions are shown as well, indicating strongly skewed distributions of arXiv article downloads, Twitter mentions and citations.}
                       
\end{figure}

\medskip In Figure~\ref{fig:scatterplots} we show the bivariate scatterplots between Twitter mentions, arXiv downloads and citations. The corresponding Pearson's correlation coefficients are shown as well. Figure~\ref{fig:tvsc} and \ref{fig:avsc} again show that Twitter mentions are correlated with citations better than arXiv downloads, which matches our results obtained from multivariate linear regression analysis. In addition, Twitter mentions are also positively correlated with arXiv downloads as is shown in Figure~\ref{fig:tvsa}, suggesting that the Twitter attention received by an article can be used to estimate its usage data, but usage, in turn, does not seem to correlated to early citations. Given the rather small sample size and the unequally distributed scatter, we performed a delete-1 observation jackknife on the Pearson's correlation coefficient between Twitter mentions and early citations (N=70). This yields a modified correlation value of 0.430 vs. the original value of 0.4516 indicating that the observed correlation is rather robust. However, dropping the top two frequently tweeted articles does reduce the correlation to 0.258 (p=0.016) implying that the observed correlation is strongest when frequently mentioned articles on Twitter are included, matching the results reported by \cite{Eysenbach_2011}.

\section{Discussion}
The ongoing move to online scholarly communication has introduced new possibilities for measuring scholarly impact. At the same time, it has become more difficult to determine which communities drive a particular form of online impact. For example, usage data, measured as volume of downloads, is generally assumed to reflect the interests and preferences of the general public, but what if the particular online service for which usage data was recorded is dedicated to serving scientists only? What if an online service for scientists increasingly becomes a tool for the general public to learn about scientific findings?  The online user communities associated with particular services may in fact overlap to various degrees as the scholarly community progressively moves online and the online public moves toward scholarly information services. Naturally, scholarly impact metrics should acknowledge this new reality.

\medskip The research presented in this paper is based on data from two services which are arguably associated with and intended for two different audiences. ArXiv.org is focused on offering scientists an online platform to publish pre-prints. Twitter is designed to serve as a micro-blogging services for the public.  In this study we did, however, not attempt to conceptualize arXiv downloads solely as scientific impact, and Twitter mentions solely as public chatter. Rather, we measured the correlation and temporal differences between these forms of responses, working under the assumption that these services naturally have overlapping and interacting user communities.

\medskip Our results, though preliminary, are highly suggestive of a strong tie between social media interest, article downloads, and even early citations. We find that Twitter mentions and arXiv downloads of scholarly articles follow two distinct temporal patterns of activity, with Twitter mentions having shorter delays and narrower time spans than arXiv downloads. We also find that volume of Twitter mention is statistically correlated with that of both downloads and ``early'' citations, i.e., citations in the scholarly record occurring less than 7 months after the publication of a preprint.

\medskip We can think of two possible explanations for these results. First, the manner in which Twitter mentions, arXiv downloads and article citations are correlated could indicate a causal relation. Scholars are increasingly exposed to social media such as Twitter, and therefore their scholarly download and citation behavior is unavoidably affected. A paper submitted to arXiv that happens to receive high levels of mentions in social media will, as a result, receive greater exposure among both the general public and scholars. As a consequence, it will receive greater levels of scholarly interest, and higher volumes of downloads and subsequent citations. Our results indeed indicate that early Twitter mentions of a paper seem to lead to more rapid and more intense download levels and subsequently higher citation levels. Second, an equally plausible, alternative explanation for our results lies in the intrinsic quality or popular appeal of different manuscripts. A manuscript of greater quality or appeal, either among the public or the scholarly community, will by virtue of this characteristic enjoy higher levels of mentions on Twitter, higher levels of downloads on arXiv, and higher levels of later citations.  As a result these indicators will seem to be correlated, and even causative of each other.

\medskip We therefore acknowledge that these observations can be the result of a number of distinct or overlapping factors which our methodology confounds and fails to distinguish. Consequently, we caution against drawing the unwarranted conclusion that these results indicate that the scholarly impact of an article can be fully determined by its social media coverage, nor that one could increase the citation rate of an article by merely tweeting about it. The fact that some correlation --- no matter how small --- was observed between social media coverage, usage, and early citations may nevertheless indicate that the scientific communication process is increasingly affected by the growing societal importance of social media. In future research we will therefore continue to focus on unraveling the potential mechanisms that tie these various factors together. These efforts might shed light on whether and how social media is becoming a component of academic and scholarly life.

\section{Materials}
\subsection{Abbreviations}
Table \ref{arxivabbrv} presents a list of the subject domain abbreviations used in this article.

\begin{table}
		\begin{center}
	\caption{\label{arxivabbrv} List of abbreviations for arXiv.org subject domains}
\begin{footnotesize}
\begin{tabular}{|l||l|}
\hline
Subject Abbr.	&   Description	\\
\hline\hline
astro-ph	&Astrophysics			\\
hep		&High Energy Physics	\\
physics	&Physics				\\
math		&Mathematics		\\
cond-mat	&Material Science	\\
cs		&Computer Science		\\
quant-ph	&Quantum Physics		\\
gr-qc		&General Relative and Quantum Cosmology	\\
nucl		&Nuclear				\\
q-bio		&Quantitative Biology\\
math-ph	&Mathematical Physics	\\
nlin		&Nonlinear Science		\\
stat		&Statistics			\\
q-fin		&Quantitative Finance	\\
\hline
\end{tabular}
\end{footnotesize}
\end{center}
\end{table}

\subsection{Data collection}
Our process of determining whether a particular arXiv article was mentioned on Twitter consists of three phases: crawling, filtering, and organization. Tweets are acquired via the Streaming API from Twitter Gardenhose, which represents roughly 10\% of the total tweets from public time line through random sampling. We collected tweets whose date and time stamp ranges from 2010-10-01 to 2011-04-30 which results in a sample of 1,959,654,862 tweets.

\medskip The goal of the data filtering process is to find all tweets that contain a URL that directly or indirectly links to any arXiv.org paper. However, determining whether a paper has or has not been mentioned on Twitter is fraught with a variety of issues, the most important of which is the prevalence of partial or shortened URLs. Twitter imposes a 140 character limit on the length of Tweets, and users therefore employ a variety of methods to replace the original article URLs  with alternative or shortened ones. Since many different shortened URLs can point to the same original URLs, we resolve all shortened URLs in our Twitter data set to determine whether any of them point to the articles in our arXiv cohort.

We distinguish between four general types of scholarly mentions in Twitter, based on whether they contain:
\begin{enumerate}
	\item a URL that directly refers to a paper published in arXiv.org. 
	\item a shortened URL that upon expansion refers to an arXiv.org paper
	\item a URL that links to a web page, e.g.~a blog posting,  which itself contains a URL that points to an arXiv.org paper.
	\item a shortened URL that links to a type (3) mention after expansion.
\end{enumerate}

\medskip In order to detect these four types of Twitter mentions, we first expand all shortened URLs in our crawled public tweets. We select the top 16 popular URL shortening services, including bit.ly, tinyurl.com, and ow.ly, and expand the shortened URLs in our collection of tweets using their respective APIs. As such, we resolved 98,377,880 short URLs, which were mostly generated by the following URL shorteners: bit.ly (61.3\%), t.co (15.2\%), fb.me (6.5\%), tinyurl.com (6.1\%) and ow.ly (4.4\%). (We acknowledge that this procedure will not identify all Twitter mentions of a given arXiv.org paper, but it will however capture most.) From the resulting set, we retain all tweets that contain the term `arXiv' and at least one URL. Next, we associate tweets to arXiv papers by extracting the arXiv ID (substrings matching `dddd.dddd') from any papers mentioned in those tweets. (Note that in the case of the third and fourth type of Twitter mention the arXiv paper ID is not explicitly shown in the tweet itself, but needs to be extracted from the web pages that the tweet in question links to.)

\section*{Acknowledgments}
This study was partly supported by NSF grant SBE \#0914939. We would like to thank staff teams of the arXiv.org service at Cornell University, the Astrophysics Data System (ADS) at the Harvard-Smithsonian Center for Astrophysics, and the bit.ly team for providing download and link data. We also thank the anonymous PLoS One reviewers for their invaluable feedback which helped us to significantly improve the manuscript.

\bibliography{refs}

\end{document}